\documentclass[aps,pre,superscriptaddress,preprint]{revtex4}
\usepackage{amsmath}
\usepackage{amsfonts}
\usepackage{amssymb}
\usepackage{graphicx}
\begin{document}
\title{Effect of ion mass on pair production in the interaction of an ultraintense laser with overdense plasmas}
\author{F. Wan}
\author{C. Lv}
\author{M. R. Jia}
\affiliation{College of Nuclear Science and Technology, Beijing Normal University, Beijing 100875, China}
\author{H. Y. Wang}
\affiliation{School of Physics Science and Technology, Anshan Normal University, Anshan 114005,
China}
\author{B. S. Xie \footnote{Corresponding author. Email address: bsxie@bnu.edu.cn}}
\affiliation{College of Nuclear Science and Technology, Beijing Normal University, Beijing 100875, China}
\affiliation{Beijing Radiation Center, Beijing 100875, China}

\date{\today}

\begin{abstract}
The effect of ion mass on pair production in the interaction of an ultraintense laser with overdense plasmas has been explored by particle-in-cell (PIC) simulation. It is found that the heavier ion mass excites the higher and broader electrostatic field, which is responsible for the enhancement of backward photon number. The pair yields are also reinforced due to the increase of head-on collision of backwards photon with incoming laser. By examining the density evolution and angle distribution of each particle species the origin of pair yields enhancement has been clarified further.
\end{abstract}
\maketitle

\section{Introduction}

The next generation of $10-100$ petawatt lasers facilities \cite{Mourou2007} are promised to be available in a near future. For example, in 2016, three $10^{24} \mathrm{W/cm^2}$ lasers will be launched as a part of the European Extreme Light Infrastructure (ELI) project \cite{ELI, Mourou2011, Fedotov2015}. These laser facilities will be capable of probing the nonlinear quantum electrodynamics (QED) physics, such as the pair production from vacuum as well the antimatter production in astrophysics environment like the pulsars and black holes \cite{Goldreich1969, Ridgers2013}. These will not only provide a deep insight of fundamental physics, but also motivate novel industrial application.

There are several different physical mechanisms to produce electron-positron pairs. With the state of the art laser, one feasible plan of initiating pair cascades is to accelerate electrons to tens of $\mathrm{MeV}$ or even $\mathrm{GeV}$ with laser beams, then impinge them to high-Z target material \cite{Shearer1973}. Pairs could be generated through trident process ($e^- + Z \rightarrow e^{-\prime} + e^- + e^+$), or through two step process \cite{ERBER1966,Anguelov1999a,Sarri2015,Vodopiyanov2015} in which the first is the photon emission from bremsstrahlung($e + Z \rightarrow \gamma + e^{\prime} + Z$) and then second is the pair production from Bethe-Heitler (BH) process ($\gamma + Z \rightarrow Z + e^- + e^+$). This type of pair cascades could be verified with moderate laser intensities $I \sim 10^{22}\mathrm{W/cm^2}$, yet the positron yields are too low for application \cite{Kirk2009}.

Another important mechanism draws researchers many interests is the spontaneous pair creation from vacuum by laser beams, i.e. the Schwinger mechanism, in which the corresponding threshold of electric field is $E_{\mathrm{crit}} \approx 1.3 \times 10^{18} \mathrm{V/m}$ \cite{Schwinger1951}. Since the conservation of energy and momentum forbids pair production in plain wave, an alternative method of standing wave by two colliding lasers could meet the requirement \cite{Bulanov2006a, Ruf2009} while the current laser facilities are not yet available for such high intensity. Some studies indicate that the pair yields are sensitive to sub-cycle information of applied laser field \cite{Hebenstreit2009,Hebenstreit2016}.

Experimentally the pair production by laser beams has been realized in the Stanford Linear Accelerator Center (SLAC) facility in 1997 \cite{Burke1997}. In this experiment, a bunch of $46.6 \mathrm{GeV}$ electrons collided with laser beams, and pairs were generated. The process could be understood as follows: at first the high energy photons are generated by nonlinear Compton scattering of laser photons with relativistic electrons and then subsequently the pairs can be created by these high energy photons interacting with laser beams through the Breit-Wheeler (BW) process \cite{Schwinger1951, Bell2008}. The difference for the pair production by BW from by BH is that the electrostatic field of nucleus in BH is replaced by an ultrastrong laser field in BW \cite{Kirk2009}. Since the interaction process in BW way is too short, the pair production yields are still very low.

Recently, a feasible plan of prolific pair production in laser plasmas interaction, also called avalanche has been suggested by several groups \cite{Bell2008, Kirk2009, Elkina2011, Ridgers2012}. By avalanche, high energy photons (hard photons) are emitted from synchrotron emission or nonlinear Compton scattering of energetic electrons, and these photons can produce pairs through BW process when they experience the ultrastrong electromagnetic field. The controlling parameters for emission and pair production are $\eta = (e\hbar /m^3 c^4)|F_{\mu \nu} p^{\nu}|$ and $\chi = e\hbar^2 /(2m^3c^4)|F^{\mu \nu}k_\nu|$ \cite{ERBER1966}, where $-e$ and $m$ is the charge and mass of electron respectively, $\hbar$ is the Plank constant, $c$ is the speed of light, $F_{\mu \nu}$ is the electromagnetic tensor, $p^{\nu}$ and $k_{\nu}$ is the four-momentum of electron and photon respectively. When $\eta$ and $\chi$ approach unity, hard photon emission and pair production are greatly enhanced \cite{Kirk2009}. Due to the high density of electrons in solid target, very huge amount of photons could be generated within the laser axial zone. And the acquired yields of positrons are large enough to be detected easily \cite{Ridgers2012}.

Many simulation researches have been performed on photon emission and pair production in laser plasmas interaction \cite{Li2014,Ji2014,Wang2015,Sagar2015,Blackburn2015,Serebryakov2015}. And some effects, for example, the laser polarization and beam configuration have been also studied to show their influences on the pair production \cite{Bashmakov2014,Gelfer2015b,Grismayer2015,Chang2015}. In particular the influence of ion mass on laser absorption and radiation are found very important \cite{Capdessus2013}. However, the effect of ion mass on pair production in $\mathrm{laser-plasma}$ interaction is still lacking of study enough, to our knowledge. This leads us to make this study in present work since we believe that the ion mass influence could be also important in the pair production.

In this paper, we use the QED-PIC code EPOCH \cite{Arber} to simulate the photon emission and pair production in laser-plasma interaction. The laser plasmas interaction is simulated by the normal PIC algorithm, and the nonlinear Compton scattering and BW pair production are simulated by the Monte Carlo (MC) method \cite{Ridgers2012}. Hard photons are treated as chargeless point particles, and they are not influenced by the electromagnetic (EM) field before pair conversion. The results show that plasmas with higher ion mass can sustain broader and stronger electrostatic field, which will drive more electrons in the laser front oscillate back and forth temporarily. This oscillation can make electrons moving anti-parallel to the incoming laser, and increase the controlling parameter of emission $\chi$, thus there will be more backwards photons. In the same way, the resulting backwards motion photons acquire large $\eta$, and can convert into pairs with a great probability. Besides, we find that one-dimensional (1D) and two-dimensional (2D) results are similar, only the relative yields are different for each case, and we attribute this to the dimension effects.

\section{Numerical simulation and results}	

In order to demonstrate the influence of ion mass, we have also chosen the plasma species like in Ref \cite{Capdessus2013}. Three kinds of plasmas with different ion species of charge +1 are examined, i.e., hydrogen plasma with protons, tritium plasma with tritium ions and immobile ion plasma with immobile ions. 		

\subsection{Simulation set up}
	
\begin{figure}
\includegraphics[width = 12cm]{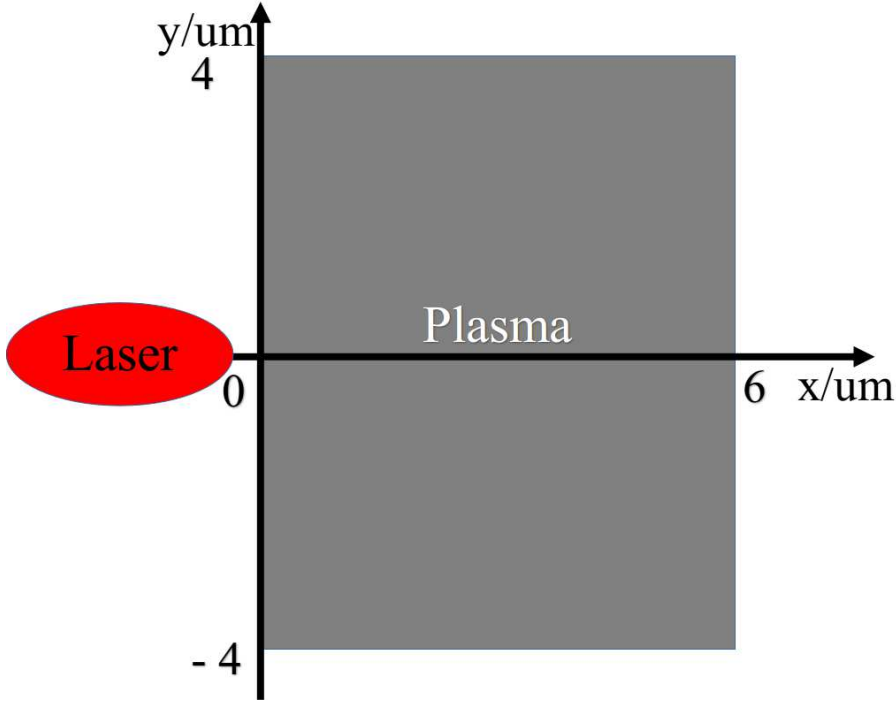}
\caption{(color online). Simulation set up}
\label{setup}
\end{figure}

We have used linearly polarized laser with a during of $30\mathrm{fs}$, constant time profile and Gaussian profile in $y$ direction with the radius of spot size $r = 1\mathrm{\mu m}$. The laser intensity is $I = 4 \times 10^{23} \mathrm{W/cm^2}$, and wavelength is chosen as $\lambda = 1\mathrm{\mu m}$. The electron and ion densities are chosen as $n_e = n_p = 100 n_c$ in all simulation and distributed uniformly from $0$ to $6 \mathrm{\mu m}$ in $x$ direction, and $-4\mathrm{\mu m}$ to $4\mathrm{\mu m}$ in $y$ direction, see Fig. \ref{setup}, where $n_c = m_e \omega^2 / 4 \pi e^2$ is the plasma critical density, $\omega$ is the laser frequency. In each direction, the simulation domain covers $8\mathrm{\mu m}$ and has been split into $800$ cells, i.e. $\Delta x = \Delta y = 0.01 \lambda$, such that the simulation is valid, i.e. the plasma frequency $\omega_{pe} \Delta x / c \ll 2$. Each cell has been filled with $120$ macro-electrons and $13$ macro-ions. The boundary conditions in each direction is set as outflow or open to reduce the effects from reflected wave or particles. For the QED part, only photons with $\epsilon_\gamma > m_e c^2$ are stored during simulation (photons with energy less than $m_e c^2$ are also emitted but are not tracked, and the recoil effects on electrons are also calculated). Besides, to minimize the computation and increase the accuracy, another group of 1 D simulations with same parameters but with $2000$ cells in $x$ direction and $400$ macro-electrons, $100$ macro-ions per cell are made. The 1D results are very similar to our rough 2D results, so we are not going to explain the results from 1D in details.

\subsection{$\gamma$ radiation and pair production}
\begin{figure}[htb]
\centering
\includegraphics[width=12cm]{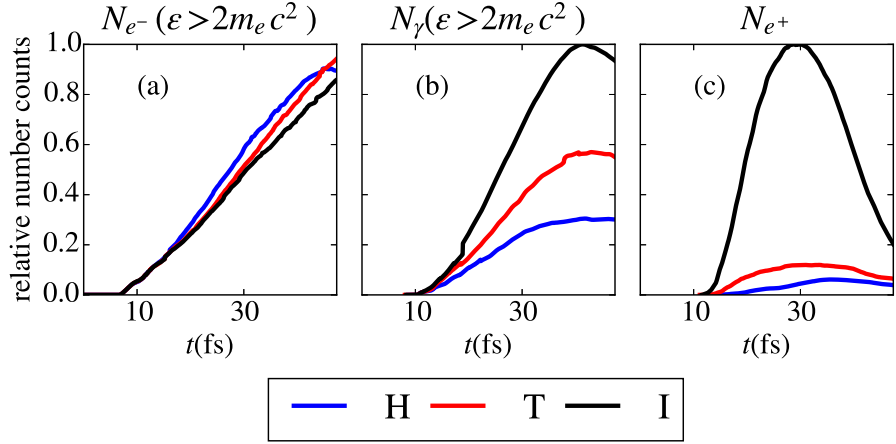}
\caption{(color online). Relative production numbers of hot electrons, $\gamma$ and $e^+$, where symbols of H (blue lines), T (red lines) and I (black lines) refer to three kind of plasmas situations with the proton, the tritium and the immobile ion, respectively.}
\label{number}
\end{figure}

\subsubsection{Electron heating}
	
\begin{figure}
\includegraphics[width=12cm]{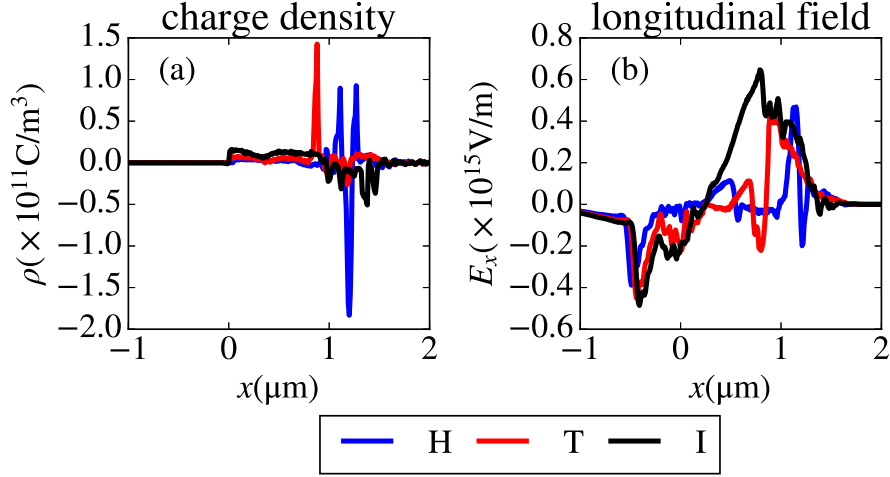}
\caption{(color online). Net charge density (a) and longitudinal field (b) at $14\mathrm{fs}$. The colored lines indicated by symbols  H, T and I are the same as in Fig.\ref{number}.}
\label{chargefield}
\end{figure}

In Fig.\ref{number}(a), we have shown the number evolution of energetic electrons, one can see that electrons heating are different in three kind of cases. Obviously the electrons is heated quicker in the case of lighter ion plasma compared to the heavier one. Theoretically owing to the linearly polarization of the laser, electrons can be heated by the oscillate Lorentz force $\mathbf{J} \times \mathbf{B}$. Yet the differences of electrons heating in three cases are caused by different longitudinal field, see Fig. \ref{chargefield}. When ion mass is heavier, their relaxation to the electronic neutral is slower, thus, a stronger and broader space charge separation field could be sustained. This force is always anti-parallel to the ponderomotive force for electrons. This could be found in Fig. \ref{chargefield}(a) where plasma with heavier ion mass leaves positive net charge density behind the laser front which builds a broader and stronger space field  \cite{Capdessus2013}. So in Fig.\ref{chargefield}(b), the narrower and weaker field is formed in the proton plasma while the broader and stronger field is formed in the immobile ion plasma.
 		
\subsubsection{$\gamma$ emission influenced by ion mass}

In Fig. \ref{number}(b), numbers of generated photons in different kinds of plasmas are plotted. The difference of photon emission in three cases is even larger than the electron heating. The emission processes initiate almost simultaneously for three kind of plasmas at $7 \mathrm{fs}$ when the laser begins to contact with the plasma, but the maximum points are different. Meanwhile, the number of electrons with $\epsilon_{e^-} > 2 m_e c^2$ and photons with $\epsilon_{\gamma} > 2m_e c^2$ are also rising and proportional to $t$. Note that in Refs. \cite{Fedotov2010a,Nerush2011,DiPiazza2012,Bashmakov2014,Narozhny2014}, the numbers of photon and $e^+$ are shown to be an exponential function of time as $N_{\gamma} \propto e^{\tau t}$ in the early interaction. However it is not so in our case. The reason may be attributed to that the interaction zone is so small that the particles with large transverse velocity can easily escape from the interaction zone in short time. So the QED-cascade is hard to happen in these configurations, which leads to finally almost a linear evolution of $N_{\gamma}$ and $N_{e^+}$ in our cases but not exponential one in cases of previous works.

\subsubsection{Pair production influenced by ion mass}

Numbers of positrons generated by BW process are plotted in Fig. \ref{number}(c). The maximum number of generated $e^+$ in tritium plasma is doubled compared with the proton case, and is even larger in the immobile ions case. In the 1D case, $N_{e^+}(T) \ge 4N_{e^+}(H)$ is found, which reminds us the importance of dimension effect. The maximum positron numbers are reached at different time, for example, the time is later for the lighter ion mass. This verifies our assumption that plasmas with heavier ions can generate much more pairs than that with lighter ions when other conditions are the same.

\begin{figure}
\centering
\includegraphics[width = 12cm]{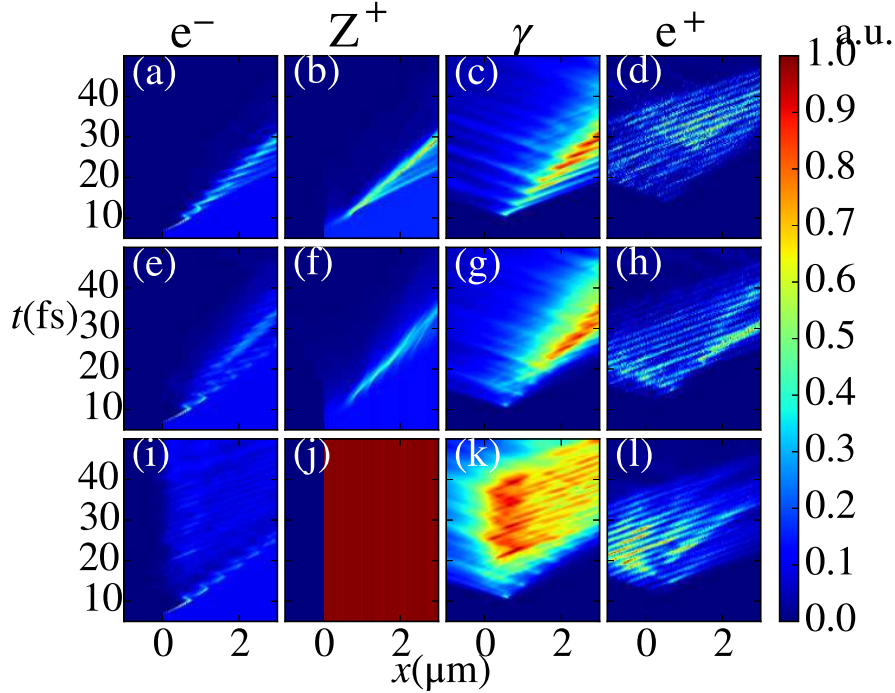}
\caption{(color online). Density evolution in $x$ direction. The rows from the top to bottom corresponds to three kinds of plasma situations with the proton, the tritium and the immobile ion, respectively.}
\label{spotx}
\end{figure}

\subsection{Analysis}

In Fig. \ref{spotx}, particle density evolution in the laser focal radius zone ($1 \mathrm{\mu m} < y < 3\mathrm{\mu m}$) along $x$ direction has been shown. The velocity of electrons piled up layer is characterized by the $v_{hb}$, hole boring velocity. Here $v_{hb}$ of electron and ion layer could be measured from the density slope in Fig. \ref{spotx}(a, e, i). However, the measured $v_{hb}$ does not coincide well with the merit of $\beta_{hb} = (\sqrt{n_c m_e /n_i m_i}a) / (1 + \sqrt{n_c m_e /n_i m_i})$ \cite{Wilks1992,Robinson2009,Qiao2012}. This is due to two factors, one is that in the linear polarized laser case, the ponderomotive force is no longer a constant pushing but with an oscillation $f_p = - (e^2/4m_e\omega^2) \nabla|E_{\mathrm{laser}}|^2(1-\mathrm{cos}2\omega t)$ \cite{Gibbon2005}. The other is that relativistic self-transparency suppresses the piston reflection and changes the hole boring to transmission induced by relativistic transparency \cite{Weng2012}. The 3rd column of Fig. \ref{spotx} contains $\gamma$ density evolution for each plasma case. One could see that after $30\mathrm{fs}$, with the increase of ion mass, more photons are generated near the plasma initial surface and propagate in the backwards direction.
		
\begin{figure}[htb]
\includegraphics[width = 12cm]{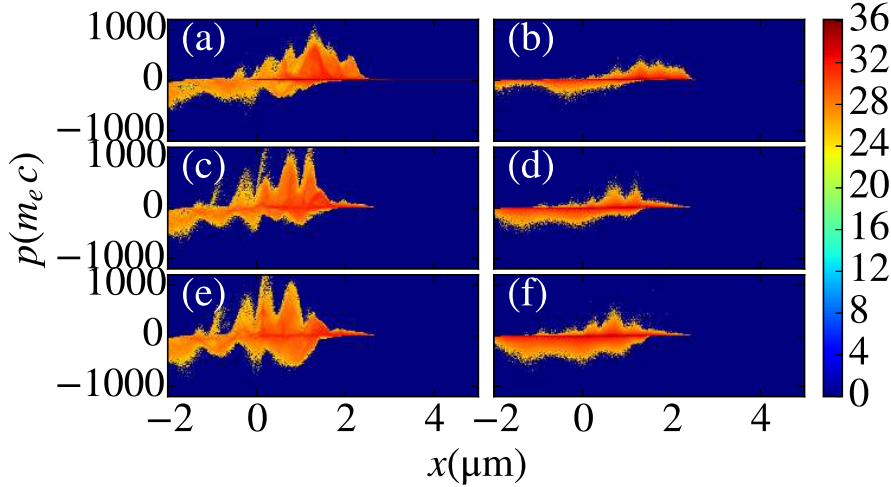}
\caption{(color online). The phase distributions of electron (left column) and photon (right column) at $14\mathrm{fs}$ in the log scale.The rows from the top to bottom corresponds to three kinds of plasma situations with the proton, the tritium and the immobile ion, respectively.}
\label{momenta}
\end{figure}	

\begin{figure}
\includegraphics[width = 12cm]{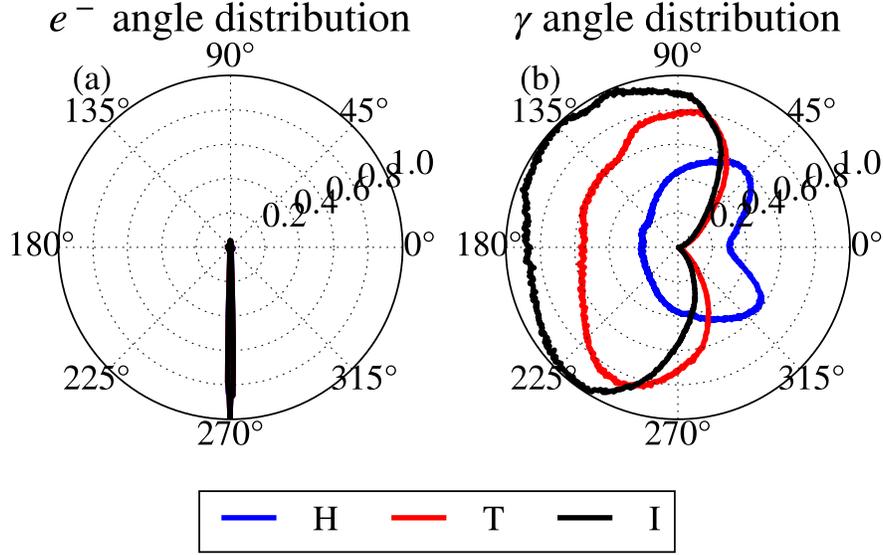}
\caption{(color online). Electron and photon angular distribution in relative counts at $14\mathrm{fs}$. The colored lines indicated by symbols H, T and I are the same as in Fig.\ref{number}.}
\label{polar}
\end{figure}

In Fig. \ref{momenta}, we have shown the phase space of electron and photon. In Fig. \ref{momenta}(a, c, e), a "train" of high energy electrons with spacing of $\delta x = \lambda_L / 2$ are driven into plasma \cite{Wilks1992}. These electrons form layers and move back and forth when they are punched by the laser. This could be attributed to the oscillatory ponderomotive force $f_p$ and longitudinal field $E_{l}$, which is verified by the oscillation frequency $2\omega$. In the Ref.\cite{Brady2012}, this kind of emission has been identified as re-injected electron synchrotron emission (RESE), i.e. emission of the driven away electrons when they are reintroduced into laser axial zone. Take the force subjected to electron layers as a function $F_{el}(x, t) = f_p - eE_{l}$, when $F_{el} > 0$, layers are accelerated in the forward direction, and when $F_{el} < 0$, layers are gradually braked or even being pulled back. In Fig. \ref{chargefield}, the net charge density and longitudinal field infer that larger ion mass could sustain larger and broader space charge field \cite{Capdessus2013}, thus drag more electrons in this oscillation. The back and forth oscillation will prolong electrons' path in strong EM field, which will generate more photons per electron. Especially, more backwards photons could be generated, which is verified in Fig. \ref{polar}(b). With heavier ion mass, lager potion of backward photons are generated \cite{Capdessus2013}, which will lead to copious high energy $\gamma$ head-on colliding with incoming laser. Therefore, more pair creation could be foreseen for heavier ion plasmas. Besides, photon emissions are also tuned by this oscillation as shown in Fig. \ref{spotx}(c, g, k). These density evolution forms lines and overlaps with electron density path in Fig. \ref{spotx}(a, e, i). Pair productions are enhanced and tuned in the same way as shown in the last column of Fig. \ref{spotx}. By the way such tuning phenomenon is absence in the circular polarized case since that in which $f_p$ is a constant.

\section{Conclusion and discussion}

In conclusion, by utilizing the PIC + MC simulation, the effect of ion mass on pair production in the interaction of an ultraintense laser with overdense plasmas has been clarified. Firstly, high $\gamma$ emission and pair production are greatly enhanced due to the increase of ion mass. The reason is that with the increase of ion mass, the excited electrostatic field is magnified. With larger excited field, more energetic electrons in the laser front can be temporarily localized or even accelerated in the backwards direction. So larger portion of electrons can collide with laser and generate backwards photons, which is most efficient for pair production.

Besides, the oscillating of ponderomotive force in the linear polarization case are also responsible for the tuning of emission and pair production. The driven back and forth motion prolongs the electron dwelling in the laser front before being accelerated away. Thus, photon emission and pair production are enhanced. Furthermore, the production of photons and pairs are very notably synchronized with the oscillation of electron layer. These results give very intuitive suggestions for pair production of laser with low Z plasmas.

\section{Acknowledgements}
This work was supported by the National
Natural Science Foundation of China (NSFC) under Grant No. 11475026. The computation was carried out at the HSCC of
the Beijing Normal University.


\begin{thebibliography}{10}

\bibitem{Mourou2007}
G. A. Mourou, C. L. Labaune, M. Dunne, N. Naumova,
and V. T. Tikhonchuk, \textit{Plasma Phys. Control. Fusion} \textbf{49},
B667 (2007).

\bibitem{ELI}
http://www.extreme-light-infrastructure.eu/.

\bibitem{Mourou2011}
G. Mourou and T. Tajima, \textit{Opt. Photonics News} \textbf{22}, 47
(2011).

\bibitem{Fedotov2015}
N. B. Narozhny and A. M. Fedotov, \textit{Physics-Uspekhi} \textbf{58},
95 (2015).

\bibitem{Goldreich1969}
P. Goldreich and W. H. Julian, \textit{Astrophys. J.} \textbf{157}, 869
(1969).

\bibitem{Ridgers2013}
C. P. Ridgers, C. S. Brady, R. Duclous, J. G. Kirk,
K. Bennett, T. D. Arber, and A. R. Bell, \textit{Phys. Plasmas}
\textbf{20}, 056701 (2013).

\bibitem{Shearer1973}
 J. W. Shearer, J. Garrison, J. Wong, and J. E. Swain,
\textit{Phys. Rev. A} \textbf{8}, 1582 (1973).

\bibitem{ERBER1966}
 T. Erber, \textit{Rev. Mod. Phys.} \textbf{38}, 626 (1966).

\bibitem{Anguelov1999a}
 V. Anguelov and H. Vankov, \textit{J. Phys. G Nucl. Part. Phys.}
\textbf{25}, 1755 (1999).

\bibitem{Sarri2015}
 G. Sarri, K. Poder, J. M. Cole, W. Schumaker, A. Di
Piazza, B. Reville, T. Dzelzainis, D. Doria, L. A. Gizzi,
G. Grittani, et al., \textit{Nat. Commun.} \textbf{6}, 6747 (2015).

\bibitem{Vodopiyanov2015}
 I. B. Vodopiyanov, J. R. Dwyer, E. S. Cramer, R. J.
Lucia, and H. K. Rassoul, \textit{J. Geophys. Res. Sp. Phys.}
\textbf{120}, 800 (2015).

\bibitem{Kirk2009}
 J. G. Kirk, A. R. Bell, and I. Arka, \textit{Plasma Phys. Control.
Fusion} \textbf{51}, 085008 (2009).

\bibitem{Schwinger1951}
 J. Schwinger, \textit{Phys. Rev.} \textbf{82}, 664 (1951).

\bibitem{Bulanov2006a}
 S. S. Bulanov, N. B. Narozhny, V. D. Mur, and V. S.
Popov, \textit{J. Exp. Theor. Phys.} \textbf{102}, 9 (2006).

\bibitem{Ruf2009}
 M. Ruf, G. R. Mocken, C. Muller, K. Z. Hatsagortsyan,
and C. H. Keitel, \textit{Phys. Rev. Lett.} \textbf{102}, 080402 (2009).

\bibitem{Hebenstreit2009}
 F. Hebenstreit, R. Alkofer, G. V. Dunne, and H. Gies,
\textit{Phys. Rev. Lett.} \textbf{102}, 150404 (2009).

\bibitem{Hebenstreit2016}
 F. Hebenstreit, \textit{Phys. Lett. Sect. B Nucl. Elem. Part.
High-Energy Phys.} \textbf{753}, 336 (2016).

\bibitem{Burke1997}
 D. Burke, R. Field, G. Horton-Smith, J. Spencer,
D. Walz, S. Berridge, W. Bugg, K. Shmakov, A. Weide-
mann, C. Bula, et al., \textit{Phys. Rev. Lett.} \textbf{79}, 1626 (1997).

\bibitem{Bell2008}
 A. R. Bell and J. G. Kirk, \textit{Phys. Rev. Lett.} \textbf{101}, 200403
(2008).

\bibitem{Elkina2011}
 N. V. Elkina, A. M. Fedotov, I. Y. Kostyukov, M. V.
Legkov, N. B. Narozhny, E. N. Nerush, and H. Ruhl,
\textit{Phys. Rev. Spec. Top. - Accel. Beams} \textbf{14}, 054401 (2011).

\bibitem{Ridgers2012}
 C. P. Ridgers, C. S. Brady, R. Duclous, J. G. Kirk,
K. Bennett, T. D. Arber, A. P. L. Robinson, and A. R.
Bell, \textit{Phys. Rev. Lett.} \textbf{108}, 165006 (2012).

\bibitem{Wang2015}
 H. Y. Wang, X. Q. Yan, and M. Zepf, \textit{Phys. Plasmas} \textbf{22},
093103 (2015).

\bibitem{Sagar2015}
 V. Sagar, S. Sengupta, and P. K. Kaw, \textit{Phys. Plasmas}
\textbf{22}, 123102 (2015).

\bibitem{Li2014}
 J. X. Li, K. Z. Hatsagortsyan, and C. H. Keitel, \textit{Phys. Rev. Lett.} \textbf{113}, 044801 (2014).


\bibitem{Ji2014}
L. L. Ji, A. Pukhov, E. N. Nerush, I. Y. Kostyukov, B. F. Shen, and K. U. Akli, \textit{Phys. Plasmas}, \textbf{21}(2), 023109 (2014).

\bibitem{Blackburn2015}
 T. G. Blackburn, \textit{Plasma Phys. Control. Fusion} \textbf{57},
075012 (2015).

\bibitem{Serebryakov2015}
 D. A. Serebryakov, E. N. Nerush, and I. Y. Kostyukov,
\textit{Phys. Plasmas} \textbf{22}, 123119 (2015).

\bibitem{Bashmakov2014}
 V. F. Bashmakov, E. N. Nerush, I. Y. Kostyukov, A. M.
Fedotov, and N. B. Narozhny, \textit{Phys. Plasmas} \textbf{21}, 013105
(2014).

\bibitem{Gelfer2015b}
 E. G. Gelfer, A. A. Mironov, A. M. Fedotov, V. F.
Bashmakov, E. N. Nerush, I. Y. Kostyukov, and N. B.
Narozhny, \textit{Phys. Rev. A} \textbf{92}, 022113 (2015).

\bibitem{Grismayer2015}
 T. Grismayer, M. Vranic, J. L. Martins, R. Fonseca, and
L. O. Silva, Arxiv 1511.07503 (2015).

\bibitem{Chang2015}
 H. X. Chang, B. Qiao, Z. Xu, X. R. Xu, C. T. Zhou,
X. Q. Yan, S. Z. Wu, M. Borghesi, M. Zepf, and X. T.
He, \textit{Phys. Rev. E} \textbf{92}, 053107 (2015).

\bibitem{Arber}
 T. D. Arber, K. Bennett, C. S. Brady, A. Lawrence-
Douglas, M. G. Ramsay, N. J. Sircombe, P. Gillies, R. G.
Evans, H. Schmitz, A. R. Bell, et al., \textit{Plasma Phys. Con-
trol. Fusion} \textbf{57}, 113001 (2015).

\bibitem{Capdessus2013}
 R. Capdessus, E. dHumieres, and V. T. Tikhonchuk,
\textit{Phys. Rev. Lett.} \textbf{110}, 215003 (2013).

\bibitem{Fedotov2010a}
 A. M. Fedotov, N. B. Narozhny, G. Mourou, and G. Korn,
\textit{Phys. Rev. Lett.} \textbf{105}, 080402 (2010).

\bibitem{Nerush2011}
 E. N. Nerush, V. F. Bashmakov, and I. Y. Kostyukov,
\textit{Phys. Plasmas} \textbf{18}, 083107 (2011).

\bibitem{DiPiazza2012}
 A. Di Piazza, C. M$
 \mathrm{\ddot{u}}$ller, K. Z. Hatsagortsyan, and C. H.
Keitel, \textit{Rev. Mod. Phys.} \textbf{84}, 1177 (2012).

\bibitem{Narozhny2014}
  N. Narozhny and A. Fedotov, \textit{Eur. Phys. J. Spec. Top.}
\textbf{223}, 1083 (2014).

\bibitem{Wilks1992}
 S. C. Wilks, W. L. Kruer, M. Tabak, and A. B. Langdon,
\textit{Phys. Rev. Lett.} \textbf{69}, 1383 (1992).

\bibitem{Robinson2009}
 A. P. L. Robinson, D.-H. Kwon, and K. Lancaster,
\textit{Plasma Phys. Control. Fusion} \textbf{51}, 095006 (2009).


\bibitem{Qiao2012}
  B. Qiao, S. Kar, M. Geissler, P. Gibbon, M. Zepf, and
M. Borghesi, \textit{Phys. Rev. Lett.} \textbf{108}, 115002 (2012).

\bibitem{Gibbon2005}
 P. Gibbon, \textit{Short Pulse Laser Interactions with Matter}
(Imperial College Press, 2005).

\bibitem{Weng2012}
 S. M. Weng, M. Murakami, P. Mulser, and Z. M. Sheng,
\textit{New J. Phys}. \textbf{14}, 063026 (2012).

\bibitem{Brady2012}
 C. S. Brady, C. P. Ridgers, T. D. Arber, A. R. Bell, and
J. G. Kirk, \textit{Phys. Rev. Lett}. \textbf{109}, 245006 (2012).

\end{thebibliography}

\end{document}